\documentclass[12pt,a4paper]{article}

\usepackage[centering,hmargin=1.7cm,vmargin=0.78in]{geometry}

\usepackage{amsmath,amssymb,titling,authblk}

\usepackage{amsmath,mathtools,amsthm,amssymb}
\usepackage{tabularx}

\usepackage{graphicx}

\usepackage{color}
\usepackage{bbold}
\usepackage{enumitem}
\usepackage{physics}
\usepackage{comment}
\usepackage{array}
\usepackage{graphics}
\usepackage{wrapfig}
\graphicspath{{figures/}}
\usepackage{bbm}
\usepackage{dsfont}
\usepackage{mathrsfs}
\usepackage{mathdots}
\usepackage{enumitem}
\usepackage{pifont}
\usepackage[table]{xcolor}
\usepackage{booktabs}
\usepackage{pifont}
\usepackage{amssymb}
\definecolor{lightred}{RGB}{243,229,231}
\definecolor{lightgreen}{RGB}{241,255,239}
\definecolor{lightblue}{RGB}{232,240,244}
\definecolor{RoyalBlue}{RGB}{65,105,225}
\definecolor{ForestGreen}{RGB}{34,139,34}   
\definecolor{Maroon}{RGB}{135,0,0}
\definecolor{myrefcolor}{rgb}{0.067,0.5,0.5}
\definecolor{myurlcolor}{rgb}{0.1,0,0.9}

\usepackage[
    breaklinks,
    pdftex,
    colorlinks=true,
    linkcolor=myrefcolor,
    citecolor=myrefcolor,
    urlcolor=myrefcolor
]{hyperref}
\usepackage[capitalize]{cleveref}

\DeclareDocumentCommand\mel{ s s m m m }
{ 
    \IfBooleanTF{#1}
    {
        \IfBooleanTF{#2}
        {\left\langle{#3}\middle\vert{#4}\middle\vert{#5}\right\rangle} 
        {\vphantom{#3#4#5}\left\langle\smash{#3}\middle\vert\smash{#4}\middle\vert\smash{#5}\right\rangle} 
    }
    {\vphantom{#4}\left\langle{#3}\middle\vert\smash{#4}\middle\vert{#5}\right\rangle} 
}

\newtheorem*{theorem*}{Theorem}

\newtheorem{definition}{Definition}

\theoremstyle{remark}

\usepackage{algorithm}
\usepackage{algpseudocode}

\newcommand{\de}{\operatorname{d}\!}

\newcommand{\be}{\begin{equation}\begin{aligned}\hspace{0pt}}
\newcommand{\ee}{\end{aligned}\end{equation}}
\newcommand{\ba}{\begin{eqnarray}}
\newcommand{\ea}{\end{eqnarray}}

\definecolor{airforceblue}{rgb}{0.36, 0.54, 0.66}

\newcommand{\bb}{\begin{equation}\begin{aligned}\hspace{0pt}}
\newcommand{\bbb}{\begin{equation*}\begin{aligned}}
\newcommand{\eb}{\end{aligned}\end{equation}}
\newcommand{\eeb}{\end{aligned}\end{equation*}}
\allowdisplaybreaks
\begin{document}

\title{From classical probability densities to quantum states: quantization of Gaussians for arbitrary orderings}




\renewcommand\Affilfont{\itshape}
\setlength{\affilsep}{1.5em}

\author[1]{Giorgio Lo Giudice\thanks{\href{mailto:giorgio99.logiudice@gmail.com}{giorgio99.logiudice@gmail.com}}}
\author[2]{Lorenzo Leone\thanks{\href{mailto:lorenzo.leone@fu-berlin.de}{lorenzo.leone@fu-berlin.de}}}
\author[1,3]{Fedele Lizzi\thanks{\href{mailto:fedele.lizzi@unina.it}{fedele.lizzi@unina.it}}}
\affil[1]{Dipartimento di Fisica ``Ettore Pancini'', Universit\`{a} di Napoli {\sl Federico~II}, Napoli, Italy}
\affil[2]{Dahlem Center for Complex Quantum Systems, Freie Universit\"at Berlin, 14195 Berlin, Germany}
\affil[3]{INFN, Sezione di Napoli, Italy}

\date{}
\maketitle

\begin{abstract}
The primary focus of this work is to investigate how the most emblematic classical probability density, namely a Gaussian, can be mapped to a valid quantum states. To explore this issue, we consider a Gaussian whose squared variance depends on a parameter $\lambda$. Specifically, depending on the value of $\lambda$, we study what happens in the classical-quantum correspondence as we change the indeterminacy of the classical particle. Furthermore, finding a correspondence between a classical state and a quantum state is not a trivial task. Quantum observables, described by Hermitian operators, do not generally commute, so a precise ordering must be introduced to resolve this ambiguity. In this work, we study two different arbitrary orderings: the first is an arbitrary ordering of the position and momentum observables; the second, which is the main focus of the present work, is an arbitrary ordering of the annihilation and creation operators. In this latter case, we find the interesting result that even a $\delta$-function, which in general has no quantum correspondence, can be mapped into a valid quantum state for a particular ordering, specifically the antinormal one (all creation operators are to the right of all annihilation operators in the product). This means that the Gaussian probability density corresponds to a valid quantum state, regardless of how localized classical particles are in phase space.
\end{abstract}

\newpage
{\em Introduction.---} 
In classical point particle mechanics a state is described by a positive, normalized to unity, positive function (in this paper we consider the one dimensional case, mainly to ease notations. There is no conceptual obstruction to go to higher arbitrary dimensions.) $\rho(q,p)$ which describes the probability density to find the particle with a certain position and velocity. Since the classical observables are functions on phase space, via the formula:
\be
\langle f\rangle_\rho=\int\de p \de q\, f(p,q) \rho(p,q)
\ee
the function $\rho$ can be seen as a map from the algebra of observables into the real numbers, whose physical meaning is the mean value of a measurement of the observable described by the real function $f$. A state therefore is a map from an algebra into numbers, which is the mathematical definition. Furthermore, we consider normalized states,  i.e.\  $\int \de q \de p \, \rho(q,p)=1$. For the sake of brevity we skip important mathematical details, like the nature of the algebra and other properties of states. 

Consider a state which describes with some uncertainty a particle in a particular point in phase space, say the origin. A natural description is a Gaussian:
\be
\rho_\lambda=\frac\lambda\pi e^{-\lambda(p^2+q^2)} \label{Gaussianpq}
\ee
This is valid for any value of $\lambda$, and the functional is well defined in the limit $\lambda\to\infty$ for which we have the Dirac $\delta$ functional:
\be
\int\de p\de q\, \delta(p,q) f(p,q)=f(0,0)
\ee
which means that the particle is in the origin without any indetermination. The generalization to an arbitrary location in phase space is straightforward. 

A linear combination of states of the kind $\alpha\rho_1+(1-\alpha)\rho_2$, for $\alpha\in[0,1]$, called a convex sum,  is still a  state. Every state described by a function (not a functional) can be written in such a way. \emph{Pure states} are those which cannot be written as a convex sum of two other states. Only the $\delta$ functionals are of this kind.

In quantum mechanics the algebra of functions of $p$ and $q$ is substituted by the algebra of their \emph{noncommuting} counterpart. Observables are self-adjoint operators, and
a state is described by \emph{density matrix}, a positive operator of trace one, whose eiegenvalues are all positive and smaller than one. Pure states are projectors, i.e.\ operators which can be written as $\hat\rho=\ket\psi\bra\psi$, for some vector in the Hilbert space. It is easy to see that the definition of pure states as states that cannot be written as convex sum of two other states, still holds.

Weyl~\cite{Weyl,suleymanov2019wigner,Zampini:2005rx,tatarskiui1983wigner} developed in the 1930s a map, called the \emph{Weyl map}, with the purpose of mapping classical probability densities to quantum states. The converse task has been extensively studied, with Wigner~\cite{Wigner} being a pioneer in this field. The Wigner function~\cite{Wigner} provides a means to associate a quantum state with a classical probability density in phase space. However, only certain quantum states can be mapped to valid classical probability densities, and in fact, the majority of these yield negative Wigner functions, resulting in quasi-probability distributions~\cite{Ferrie_2008,Ferrie_2011}. Following Wigner's work, this line of research has proven to be highly insightful, as the negativity of the Wigner function is regarded as an indicator of quantumness~\cite{Veitch_2012}, preventing the efficient classical simulation of continuous variable states~\cite{Mari_2012} and enabling the onset of quantum advantage~\cite{Gard_2015,doi:10.1126/science.abe8770,Lund_2017}.  These processes of mapping classical states to quantum states and vice-versa are based on a pair of \emph{Quantizer} and \emph{Dequantizer} operators, which we will describe in detail below.

The Weyl map associates, for example, to the classical function $pq$ the operator $\frac12(\hat p \hat q +\hat q \hat p)$. This is not the only quantization map possible, for example one can choose to put all the $\hat p$'s on the left. 

However, this is not the end of the story. Focusing on the creation and annihilation operators Cahill and Glauber~\cite{Cahill:1969it}, in the context of their studies on quantization and coherent states, introduced alternative quantizations, using the isomorphism between phase space and the complex plane, parametrized by an \textit{asymmetry parameter} $s$. Depending on the value of $s$, different orderings are achieved: for $s=0$ we have the symmetric ordering, while for $s=1$  (resp.~$s=-1$) we have  the  normal (resp.\ antinormal) ordering~\cite{PhysRevA.90.013810,ivan2012measure,tan2020negativity, de2019measuring, berra2020coherent, berra2020star}.

The aim of this paper is to investigate in which cases the quantization of the classical Gaussian state~\eqref{Gaussianpq} gives rise to a viable quantum density matrix, and to do this for arbitrary ordering of the operators $\hat p, \hat q$ or the creation and annihilation operators. For a review of quantum phase space distributions one can see~\cite{Lee1995}.

The issue may also be relevant for the quantization of gravity giving rise to a quantum spacetime and a noncommutative geometry~\cite{Kempf:1994su, Brandenberger:2002nq, Chamseddine_2010, Devastato:2019grb, Huggett:2020kok}. The issue in this case is asking in which case we can have a viable localised state when quantizing spacetime and it leads to the locality of the theory.

{\em Technical preliminaries.---} 
Throughout this work, we will focus on the phase space $\mathbb{R}^{2}$ of a single particle with position $q$ and momentum $p$, as well as the Hilbert space $\mathcal{H}$ of square-integrable functions, where position and momentum operators will be denoted as $\hat{q},\hat{p}$, respectively. We work in natural units where $\hbar=1$. 


Since we are interested in finding a correspondence between a classical state, which is a Gaussian, and a quantum state, let us briefly introduce the concept of \textit{Gaussian} states~\cite{WANG_2007,Weedbrook_2012,adesso2014continuous,genoni2016conditional} in quantum mechanics. These are an interesting class of quantum states, as they can be efficiently classically simulated~\cite{Mari_2012} and learned~\cite{mele2024learningquantumstatescontinuous}. As the name suggests, analogous to classical Gaussian distributions, these are density operators uniquely characterized by their first and second moments. Further properties of Gaussian states have been included in the appendix~\cref{Gaussian}, as they provide useful context for understanding certain comments made in the main discussion, but do not represent the core focus of the primary analysis.

For the reminder, it is useful to define $p$ and $q$ on a quantum plane defining 
\be
z\coloneqq \frac{1}{\sqrt{2}}({q}+ i{p})\ ; \ \bar z\coloneqq\frac{1}{\sqrt{2}}({q}- i{p}) \label{defzbaz}
\ee
and their quantum counterpart, the
creation and annihilation operators
\be
a\coloneqq \frac{1}{\sqrt{2}}(\hat{q}+ i\hat{p}) \ ; \ a^{\dag}\coloneq\frac{1}{\sqrt{2}}(\hat{q}- i\hat{p})
\ee
operators (where we omit the hat on $a,a^\dag$ to lighten notations). We also introduce the \textit{Fock basis} $\{\ket{k}\}_{k=0}^{\infty}$, i.e.\  eigenvectors of the photon number operator $\hat{N}=a^{\dag}a$ with eigenvalue $\hat{N}\ket{k}=k\ket{k}$.

Let us introduce the concepts of quantization map, quantizer, and dequantizer. Since we use the complex coordinate $z$ in the most significant part of this work (the Cahill-Glauber quantization), we will introduce these quantities using this coordinate. \begin{definition}[quantization map and quantizer]
      A \emph{quantization map} $\hat \Omega$ uses an operator called the \emph{quantizer} which appear in the expression
\be
\hat\Omega(f)=\int \de^2 z f(z) \hat D(z)
\ee 
where $\int\de^2 z=\int\de z\int\de\bar{z}$.
\end{definition}

\begin{definition} [dequantization map and dequantizer] The inverse (dequantization) map is obtained via another $z$ dependent operator called the \emph{dequantizer} $\hat U(z)$:
\be
\Omega^{-1}(\hat A)=\Tr \hat A \hat U(z)
\ee
\end{definition}
For the two maps to be one the inverse of the other it must be:
\be
\Tr\hat D(z)\hat U(z')=\delta(z-z')
\ee
We also require that the map is an isometry, in the sense that an $L^1$ function it associated a trace class operator, whose trace equals the integral of the original functions.

{\em Weyl quantization.---} The main focus of this work is to explore how the classical Gaussian state~\eqref{Gaussianpq} can be associated with a quantum state, i.e.\ positive trace-one operators $\hat{\rho}(q,p) \in \mathcal{H}$. Given the inherent noncommutativity of position and momentum operators in quantum mechanics, it is necessary to find a way to map the commutative product $qp$ onto the corresponding operators $\hat{q},\hat{p}$. This mapping will henceforth be referred to as a \textit{quantization map}. The issue is further complicated by the fact that noncommutativity introduces ambiguity in the ordering of these operators, and multiple orderings are possible. Therefore, we begin by considering an arbitrary ordering of $q$ and $p$ using the following quantization map:
\be
pq\mapsto \frac{1+\gamma}{2}\hat{q}\hat{p}+\frac{1-\gamma}{2}\hat{p}\hat{q}\label{eq:ordering}
\ee
with $-1\le \gamma\le 1$. More rigorously, following Weyl~\cite{Weyl, Moyal, Groenewold:1946kp, Curtright:2011vw}, we can introduce the quantization map $\hat{\Omega}_{\gamma}$.
\begin{definition}[Weyl quantization] \label{def:quantizer} Let $-1\le \gamma\le 1$. The Weyl map $\hat{\Omega}_{\gamma}(\cdot)\,:L_2(\mathbb{R}^{2})\rightarrow \mathcal{B}(\mathcal{H})$, is defined as
\be
\hat{\Omega}_{\gamma}(f)=\int\de x\de y \tilde{f}(x,y)e^{-i(x\hat{q}+y\hat{p})}e^{\frac{i}{2}\gamma xy} \label{weyl quantization}
\ee
where $\tilde{f}(x,y)$ is the Fourier transform of $f(q,p)\in L_{2}(\mathbb{R}^2)$. 

\end{definition}
It is easy to show that the action of the map $\hat{\Omega}_{\gamma}$ returns the ordering in Eq.~\eqref{eq:ordering}. 

Notice that in~\eqref{weyl quantization} only for the symmetric ordering , i.e.\ $(\gamma=0)$,  we have an Hermitian operator, which is one of the necessary conditions to have a quantum state, as the expectation values of Hermitian operators must be real. Consider the adjoint of expression \eqref{weyl quantization} and making the substitution $\xi \rightarrow -\xi$ and $\eta \rightarrow -\eta$.

\begin{equation} \hat{\Omega}^{\dagger}_{\gamma}(f) = \int \de{(-\xi)} \de{(-\eta)} \tilde{f^{*}}(-\xi, -\eta) e^{-i(\xi\hat{q} + \eta\hat{p})} e^{\frac{-1}{2}(i\gamma\xi\eta)}. \end{equation}
Given that $f(q,p)$ is real, we have $\tilde{f^{*}}(-\xi,-\eta)=\tilde{f}(\xi,\eta)$ (in this case, we used "*" to describe the complex conjugate, to lighten the notation) known as the reality condition. This leads to:
\begin{equation} \hat{\Omega}^{\dagger}_{\gamma}(f) = \int \de{\xi} d{\eta} \tilde{f}(\xi, \eta) e^{-i(\xi\hat{q} + \eta\hat{p})} e^{\frac{-1}{2}(i\gamma\xi\eta)}. \end{equation}
Comparing this result with the original map \eqref{weyl quantization}, we observe that the adjoint of the operator generally does not coincide with the operator itself, implying that the operator is not hermitian for a generic $\gamma$. Since this discrepancy is caused by the phase term $e^{\frac{-1}{2}(i\gamma\xi\eta)}$, when it disappears (i.e.\ $\gamma=0$), the operator $\hat{\Omega}_{\gamma}(f)$ is hermitian.
Therefore for the reminder of the section we make exclusively use of the symmetric ordering and denote $\hat{\Omega}(\cdot)$ the quantizer corresponding to $\gamma=0$.

As anticipated we now focus on the most emblematic classical probability density, i.e.\ a Gaussian $\rho_{\lambda}(q,p)$ defined in~\eqref{Gaussianpq},  with mean $0$ and square-variance $(2\lambda)^{-1}$. This case has been studied also in Eq.~\cite{derezinski2020,cahen2023complexweylsymbolsmetaplectic}.

Eq.~\eqref{Gaussianpq} describes the state of a classical particle sitting at the origin of phase space, i.e.\ having expected position and momentum $0$, with concentration governed by the width of the Gaussian $(2\lambda)^{-1}$. Intuitively, although this probability distribution may have a quantum analogue, the mapping might fail depending on the width of the Gaussian. In the limit of null variance, one obtains $\lim_{\lambda\rightarrow\infty}\rho_{\lambda} = \delta(p,q)$, a peaked $\delta$-function having no \textit{a-priori} physical correspondence in the quantum world. In the following proposition, we prove this intuition to be correct, showing the existence of a critical value $\lambda_c$ for which the quantizer gives rise to a valid quantum state.

Directly using the expression of the quantization map in~\cref{def:quantizer}, we compute $\hat{\Omega}(\rho)_{nm}$ in the Fock basis, resulting in
\be
\hat{\Omega}(\rho)_{nm}=\begin{cases}
    \frac{2\lambda}{1+\lambda}\left(\frac{1-\lambda}{1+\lambda}\right)^n \quad &n=m\\
    0\quad &n\neq m\\
\end{cases}
\ee
i.e.\ a state diagonal in the Fock basis. The detailed calculation is found in~\cref{App:proofprop1}. From the above expression, it is immediate to see that the eigenvalue are all positive (and less than $1$) for $\lambda<\lambda_c=1$, $\hat{\Omega}(\rho)_{00}>1$ for $\lambda>\lambda_c$, resulting into a non-positive quantum state, while $\hat{\Omega}(\rho)=\ketbra{0}{0}$ for $\lambda=\lambda_c$. This means that the quantization of the Gaussian in Eq.~\eqref{Gaussianpq} corresponds to a valid quantum state only if its squared variance $(2\lambda)^{-1}$ is larger than $1/2$. In particular, it corresponds to a pure quantum state only if the squared variance is exactly equal to $1/2$. This fact hints at the possibility that the Weyl quantization maps probability distributions to positive operators if and only if a classical analogue of the Heisenberg uncertainty principle is fulfilled. For the Gaussian in Eq.~\eqref{Gaussianpq}, one has $\mathbb{E}_{\rho_{\lambda}}[q^2]=\mathbb{E}_{\rho_{\lambda}}[p^2]=(2\lambda)^{-1}$, whereas $\mathbb{E}_{\rho_{\lambda}}[q]=\mathbb{E}_{\rho_{\lambda}}[p]=0$ (with $\mathbb{E}$ denoting the expectation value) which would return a classical analogue of the Heisenberg uncertainty principle $\Delta_q \Delta_p \ge 1/2$, where $\Delta_{q}^2\coloneqq\mathbb{E}[q^2]-\mathbb{E}[q]^2$ and the same for $\Delta_p$.

It is interesting to note that, since for $\lambda > 1$ we do not obtain a valid quantum state, this implies that with Weyl quantization, we cannot map the $\delta$-function to a quantum state. In other words, we cannot ``squeeze''  the Gaussian excessively, there exists a minimum value for the squared variance that corresponds to a quantum state: $1/2$.

Let us also show that Heisenberg's uncertainty is at work. As stated above,  for $\lambda>1$ the map still associates a Hermitian operator, but it cannot be considered a possible quantum state. Let us calculate, in the case in which $\rho$ is  density matrix, the quantities $\Tr \hat q \,\hat \Omega(\rho)$ and $\Tr \hat q^2\, \hat \Omega(\rho)$ to verify Heisenberg's principle. The calculation is not difficult expressing $\hat q=\frac1{\sqrt 2}(a+a^\dag)$. Since the matrix is diagonal, only the terms with an equal number of $a$ and $a^\dag$ survive. Therefore immediately one sees that $\Tr \hat q \, \hat \Omega(\rho)=0$. Which is a consequence of the fact that we centered the Gaussian in the origin.

For the square we have to use the fact that 
 \be
 \hat q^2=\frac12(\hat a+\hat a^\dag)^2=\frac12\left(a^2+{a^\dag}^2+aa^\dag+a^\dag a\right)=\frac12\left(a^2+{a^\dag}^2+2N+1\right)
 \ee
 and therefore
 \be
\Tr \hat q^2\, \hat \Omega(\rho)=\frac12\frac{2\lambda}{1+\lambda}\sum_{n=0}^\infty(2n+1)\left(\frac{1-\lambda}{1+\lambda}\right)^n=\frac{1}{2 \lambda}
\ee
since the uncertainty for $\hat p$ is the same (the system is symmetric in $p$ and $q$), this shows that the Heisenberg principle is respected for the values of $\lambda$ for which $\hat{\Omega}(\rho)$ is a viable density matrix. For $\lambda=1$ the uncertainty is minimal (1/2 in units for which $\hbar=1$), and this is correct since in this case the matrix density corresponds to a coherent state.

It is noteworthy that the Gaussian can be interpreted as the Wigner function of a thermal state at inverse temperature $\beta$ with Hamiltonian $\hat{H}=\hat{q}^2+\hat{p}^2$. It is known (see for example~ \cite{BUCCO}) that in this case, the Wigner function is $e^{-\frac{q^2+p^2}{2\beta+1}}$. Thus, we see that $\beta = 0$ corresponds to $\lambda = 1$ in \eqref{Gaussianpq}, representing a pure state; for $\beta > 0$, we have $\lambda < 1$, indicating a mixed state; and for $\beta < 0$, we find $\lambda > 1$, which does not yield a physical state. This serves as a useful sanity check: for a state with a negative temperature, i.e.\ a non-physical state, we end up in a scenario where there is no correspondence between a classical state and a quantum one.

All our comments were based on the Guassian density probability~\eqref{Gaussianpq}. There is an interesting property for a general probability distribution: let $\rho(q,p)$ be a phase-space probability distribution with $\Delta_q$, $\Delta_p$ variances of $q,p$ respectively. If $\Delta_{q}^2+\Delta_{p}^2<1$, then $\hat{\Omega}(\rho)$ is not a positive operator and $\Delta_q\Delta_p< \frac{1}{2}$. Moreover, if the probability density $\rho(q,p)$ obeys $\Delta_q=\Delta_p$, then it also holds: if $\Delta_q\Delta_p< \frac{1}{2}$ then $\hat{\Omega}(\rho)$ is not a positive operator, hence not a valid quantum state.
The full calculation is in~\cref{App:proofth1}. Notice also that if the probability density is such that the variances of $p$ and $q$ are the same, the condition $\Delta_q^2+\Delta_p^2\ge 1$ is equivalent to $\Delta_q\Delta_p\ge\frac{1}{2}$. It means that the symmetric Weyl quantization yields physical operators in the Hilbert space only for certain classical probability densities. Specifically, for symmetric probability densities, a necessary condition is given by the classical version of the Heisenberg uncertainty principle, which asserts that a probability density may correspond to valid quantum states only if $\Delta q\Delta p \ge \frac{1}{2}$, meaning the classical uncertainty must be sufficiently large. This perfectly alings with what we showed for the Gaussian \eqref{Gaussianpq}.



{\em Cahill-Glauber quantization.---}
We now come to the central point of the paper and shift our focus to the annihilation and creation operators and their orderings.

The phase space $\mathbb{R}^2$ is isomorphic to the complex plane $\mathbb{C}$, with the coordinate $z$ defined in~\cref{defzbaz}.  Any classical probability density can be expressed as a function of $z$ and its complex conjugate $\bar{z}$. A direct quantum analogue of $z$ and $\bar{z}$ can be viewed as the annihilation and creation operators $a$ and $a^{\dag}$, and the correspondence $z \mapsto a\, (\bar{z}\mapsto a^{\dag})$ is a valid quantization. As before, however, the situation is more subtle, as multiple orderings of $a$ and $a^{\dag}$ are possible. We can therefore consider an arbitrary ordering, parameterized by $s \in [-1,1]$, as follows
\be
z\bar{z}\mapsto  \frac{1+s}{2}a^{\dag}a+\frac{1-s}{2}aa^{\dag}\label{sordering}\,,
\ee
and investigate the validity of these orderings by requiring certain fundamental physical principles to hold, following the approach outlined in the previous section. Eq.~\eqref{sordering}  confirms that for $s=0$ we have the symmetric ordering, while for $s=1$  (resp.~$s=-1$) we have the  normal (resp.\ antinormal) ordering.

\begin{figure}
    \centering
    \includegraphics[scale=0.55]{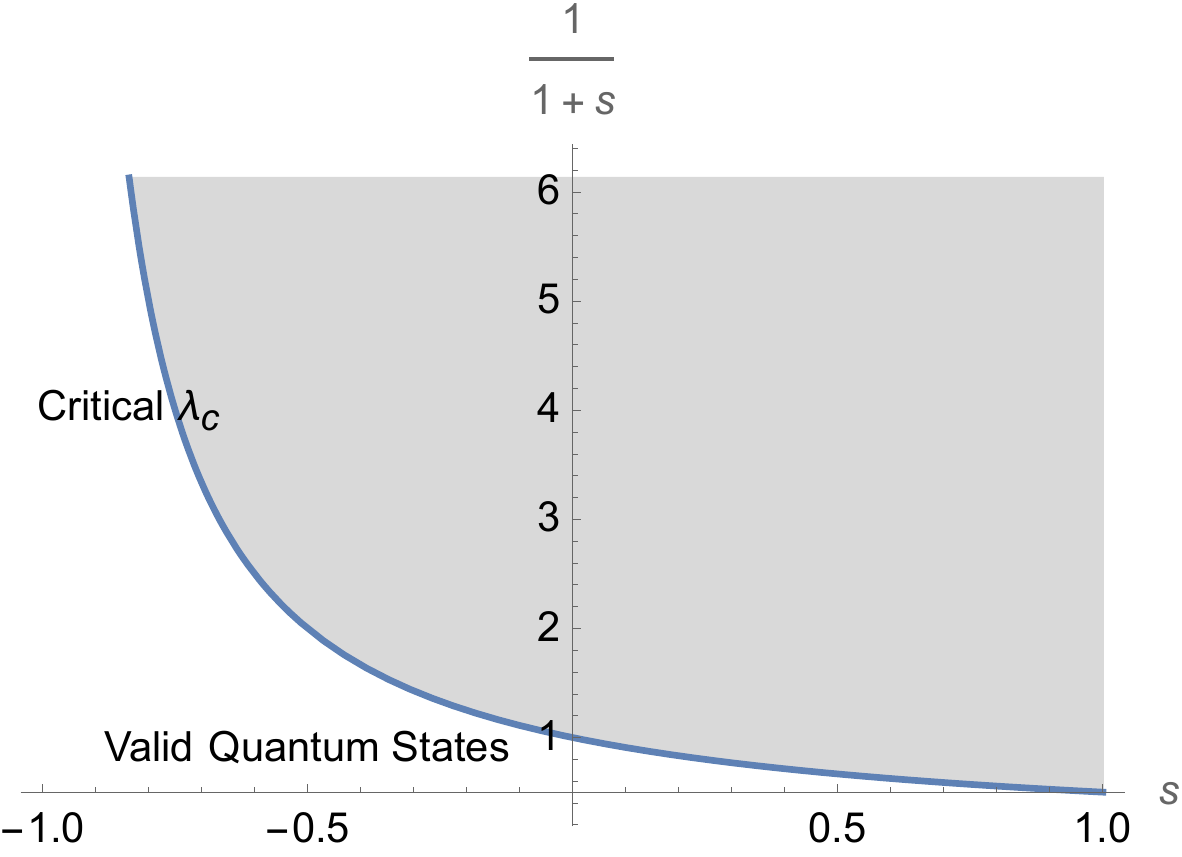}
    \caption{\textsl{
    The curve represents the critical line of the (inverse square) variance $\lambda_{c}(s)=(1+s)^{-1}$. In particular, Gaussian probability densities with $\lambda>\lambda_c(s)$ get mapped from the quantization map $\hat{\Omega}_s$ to non-positive (yet hermitian and normalized) operators, here represented by the gray region. Conversely, for $\lambda<\lambda_{c}(s)$ they get mapped to positive operators, i.e.\  valid quantum states. }}
    \label{fig:enter-label}
\end{figure}

Following Cahill and Glauber~\cite{Cahill:1969it}, we introduce the corresponding quantization map $\hat{\Omega}_{s}(f)$.
\begin{definition}[Cahill-Glauber quantization]\label{def:glauberquantizer} Let $-1\le s\le 1$. The Cahill-Glauber map $\hat{\Omega}_s(\cdot)$ is defined as
\be
\hat{\Omega}_{s}(f)\coloneqq\int\de^2\xi \tilde{f}(\xi)e^{\xi a^{\dag}}e^{-\bar{\xi}a}e^{-\frac{1-s}{2}|\xi|^2} \label{s quantization}
\ee
where $\int\de^2\xi\coloneqq\int\de\xi\int\de\bar{\xi}$ and $\tilde{f}(\xi)$ is the Fourier transform of $f(z)\in\mathbb{L}^2(\mathbb{C})$.  
\end{definition}
It is easy to see that the action of the map $\hat{\Omega}_s$ gives rise to the orderings in Eq.~\eqref{sordering}.  Unlike the previous case, where Weyl quantization yielded Hermitian operators only for $\gamma=0$, we have that the quantization map in~\cref{def:glauberquantizer} always returns a Hermitian operator for all $s\in[-1,1]$ and all real functions $f(z)=\bar{f}(z)$. Consider the adjoint of expression \eqref{s quantization} and making the substitution $\xi \rightarrow -\xi$ 
\ba
\hat{\Omega}^{\dagger}_{s}(f)&=&\int\de^2\xi \tilde{f}^*(-\xi)\Bigl(e^{-\xi a^{\dag}}e^{\bar{\xi}a}\Bigr)^{\dagger}e^{-\frac{1-s}{2}|\xi|^2}\nonumber\\
&=& \int\de^2\xi \tilde{f}(\xi)e^{\xi a^{\dag}}e^{-\bar{\xi}a}e^{-\frac{1-s}{2}|\xi|^2}=\hat{\Omega}_{s}(f)
\ea
We have proven that $\hat{\Omega}_{s}(f)$ is Hermitian for all real functions, since we again used the reality condition $\tilde{f}^*(-\xi) = \tilde{f}(\xi)$ (in this case, we used "*" to describe the complex conjugate, to lighten the notation). This holds for all real values of the parameter $s$, in particular for $s \in [-1,1]$. 
We do not limit the domain of the quantizer $\hat{\Omega}_s$ to any specific value of $s$, since, in principle, any ordering of $a$ and $a^{\dag}$ yields a valid quantum state, parametrized by the asymmetry parameter $s$. To gain intuition, we proceed with the explicit quantization of the Gaussian probability distribution parametrized by $\lambda$ in Eq.~\eqref{Gaussianpq}, which in the complex coordinates reads
\be\label{eq:complexgaussian}
\rho_{\lambda}(z)=\frac{\lambda}{\pi}e^{-2\lambda|z|^2}\,.
\ee
We thus quantize $\rho_{\lambda}(z)$ as a function of $s$, showing the existence of a critical \textit{line} $\lambda_c(s)$ for which the quantization returns a non-positive operator.
    \be
    \rho_{nm}=
\begin{cases}
\frac{2\lambda}{1+(1-s)\lambda}\left(\frac{1-(1+s)\lambda}{1+(1-s)\lambda}\right)^n\quad & n=m\\
0 & n\neq m
\end{cases}
    \ee
We perfomed this calculation in the Fock basis. The detail can be found in \cref{App:proofofprop2} . Moreover, for $s\geq-1$,  $\lambda_{c}(s)\coloneqq(1+s)^{-1}$ corresponds to a critical line, dependent on the parameter $s$. In particular,  for every $\lambda\le\lambda_{c}(s)$, $\hat{\Omega}_s(\rho_{\lambda})$ corresponds to a valid quantum state, while for $\lambda>\lambda_{c}(s)$ the corresponding operator is non-positive. 
We refer to~\cref{fig:enter-label} 
for a visual representation of the findings, which shows the critical line $\lambda_c(s)$ distinguishing physical from non-physical quantization. Interestingly, and somewhat counterintuitively, for $s = -1$, corresponding to the \textit{antinormal} ordering, even in the limit of zero variance — which represents a deterministically localized particle in phase space — the result is a physical operator in the Hilbert space:
\be\label{eq:delta}
\hat{\Omega}_{-1}(\delta(z))=\ketbra{0}{0}
\ee
i.e.\ the Dirac-delta gets mapped to the vacuum state in the Fock basis. This is a major difference compared to what we showed for the Weyl ordering, where the Dirac delta was not mapped to a quantum state. This implies that, with antinormal ordering, we do not face the issue of ``over-squeezing'' the Gaussian.

We note, in passing, that formally for $s<-1$ all Gaussian states, including the limit in which they are $\delta$, are viable quantum states.

Repeating the same analysis we performed with the Weyl quantization map regarding the temperature, we expect a shift in the temperature due to the presence of the asymmetry parameter $s$ (while in the Weyl analysis, we set $\gamma=0$ to ensure a Hermitian operator). Moreover, we must compare the Fourier transform of the Gaussian in \eqref{eq:complexgaussian} with that of the Wigner function $e^{-\frac{q^2+p^2}{2\beta+1}} = e^{-\frac{2|z|^2}{2\beta+1}}$, since the quantizer is defined through the Fourier transform. The full calculation is in appendix~\ref{App:temperature}. It turns out that $\beta=(1+s)^{-1}=\lambda_c(s)$. This means that for $\beta \geq 0$, we have $\lambda \le \lambda_c(s)$, corresponding to a valid quantum state, while for $\beta < 0$, which indicates a negative temperature, we have $\lambda > \lambda_c(s)$, corresponding to a non-physical state.

The Heisenberg principle is again similar to what we showed in the Weyl case:
\be
\Tr \hat q^2 \hat \Omega_s(\rho(z))=\frac12\frac{2\lambda}{1+(1-s)\lambda}\sum_{n=0}^\infty(2n+1)\left(\frac{1-(1+s)\lambda}{1+(1-s)\lambda}\right)^n=\frac{1-\lambda s}{2 \lambda}
\ee
and the inequality is respected for for the values of $\lambda<\lambda_c$ for which the operator can be interpreted as a density matrix.  

Similar in spirit to what we did with the Weyl quantization, we may comment about a general probability density: let $\rho(q,p)$ be a phase space probability density with $\Delta_q,\Delta_p$ variances of $q,p$ respectively. If $\Delta_{q}^2+\Delta_{p}^2<1+s$, then $\hat{\Omega}_{s}(\rho)$ is not a valid quantum state and $\Delta_q\Delta_p<\frac{1+s}{2}$. Moreover, if the probability density obeys $\Delta_q=\Delta_p$, then it holds that if $\Delta_{q}\Delta_{p}<\frac{1+s}{2}$, then $\hat{\Omega}_{s}(\rho)$ is not a valid quantum state. The calculation is in Appendix~\ref{App:proofth3}. 
The above property presents a classical version of the uncertainty principle that varies with the parameter $s$. Specifically, for the antinormal ordering ($s = -1$), the above property becomes completely irrelevant, meaning that, at least in principle, any classical probability distribution can have a quantum analogue, which is consistent with the fact that even the Dirac delta function can be quantized for this ordering, cfr.\ Eq.~\eqref{eq:delta}. This suggests that the ordering of the creation and annihilation operators fundamentally determines whether a valid quantum state can be obtained, regardless of how localized the classical probability density is. Interestingly, the antinormal ordering consistently yields a valid positive quantum state, implying a hierarchy of orderings that is dictated by the physicality of the operators.

{\em Summry and conclusions.---} In this paper, we analyzed when a Gaussian probability density can be mapped to a valid quantum state for various orderings. First, we explored the effects of implementing an arbitrary ordering of position and momentum operators. In this case, we obtain a Hermitian operator only with symmetric ordering, which imposes a constraint on the possible values of $\gamma$, determining the specific ordering of $\hat{q}$ and $\hat{p}$. Additionally, we found that for a valid quantum state, a particle cannot be too sharply localized. More precisely, we identify a critical value $\lambda_c = 1$, where $(2\lambda)^{-1}$ represents the squared variance of the Gaussian. A quantum state is possible only for $\lambda \geq \lambda_c$. This implies, for example, that we cannot map a $\delta$-function to a valid quantum state with symmetric ordering. This result aligns with the Heisenberg uncertainty principle, which is satisfied only for states with $\lambda \geq \lambda_c$. 

The central point of the present work has been the case of an arbitrary ordering of the annihilation and creation operators. In this situation we always find an hermitian operator, hence we do not have a limitation on the values for the parameter $s$. Similarly to the previous case, we find a critical value, actually a critical line $\lambda_c$, which depends on $s$. This critical line is such that only for $\lambda \geq \lambda_c$ we get a quantum state, as shown by~\cref{fig:enter-label}. The most interesting result is found for the antinormal ordering, i.e.\ $s=-1$, corresponding to a critical value $\lambda_c$ which diverges. In this situation we find that even when the variance approaches zero, indicating a particle that is perfectly localized in phase space, the outcome remains a valid operator within the Hilbert space. This means that even sharply peaked classical probability densities can correspond to valid quantum states.

\subsubsection*{Acknowledgments.} F.A.~Mele, S.F.E.~Oliviero and P.~Vitale are thanked for comments and inspiring discussions. F.L.~acknowledges support from the INFN Iniziativa Specifica GeoSymQFT, and from Grants No.~PID2019–105614 GB-C21 and No.~2017-SGR-929, support by ICSC — Centro Nazionale di Ricerca in High Performance Computing, Big Data
and Quantum Computing, funded by European Union — NextGenerationEU. 
L.L.\ is funded through the Munich Quantum Valley project (MQV-K8) by Bayerisches Staatsministerium für Wissenschaft und Kunst. 
This publication is based upon work from COST Action CaLISTA CA21109 supported by COST (European Cooperation in Science and Technology).


\appendix
\section{Gaussian states }\label{Gaussian}
Here we introduce some properties of Gaussian states used in the main text. Let us define the operator-valued vector $\hat{R}=(\hat{q},\hat{p})$. The \textit{first moment} of $\rho$ is defined as $m(\rho)\coloneqq\tr \small(\hat{R}\rho\small)$, and the \textit{covariance matrix} of $\rho$ is defined as 
\[
V(\rho)\coloneqq \tr(\{(\hat{R}-m(\rho)),(\hat{R}-m(\rho))^{T}\})\,.
\]
where $\{\cdot,\cdot\}$ is the anticommutator. The covariance matrix $V(\rho)$ is a symplectic matrix, satisfying $V(\rho)\Omega V^{T}(\rho)=\Omega$, where $\Omega$ is the usual symplectic form, and it also satisfies 
\be
V(\rho)+i\Omega\geq 0\,.\label{eq:positivity}
\ee
A Gaussian state $\sigma$ is uniquely identified by its first moment $m(\rho)$ and \textit{covariance matrix} $V(\rho)$. The converse also holds: given a valid covariance matrix and a vector $m \in \mathbb{R}^2$, there is a unique Gaussian state associated with them. In particular, a symplectic matrix is a valid covariance matrix if it fulfills Eq.~\eqref{eq:positivity}. 
Given this result, it is natural to define the \textit{Gaussification} $G(\rho)$ of a given state $\rho$, that is, the Gaussian state $G(\rho)$ identified by the first moment $m(\rho)$ and the covariance matrix $V(\rho)$ corresponding to $\rho$. 

\section{Weyl quantization}

\subsection{Weyl ordering}\label{App:proofprop1}
To determine the quantization of the Gaussian~\eqref{Gaussianpq}, we need to evaluate:
\be
\rho_{nm}=\bra n\hat\Omega(f)\ket m=\int \de x \de y \bra n e^{-i y \hat{q}-i x \hat{p}}\tilde \rho_\lambda(x,y) \ket m \label{Omega0}
\ee
where
\be
\tilde\rho_\lambda(x,y)=\frac1{2\pi}e^{-\frac{x^2+y^2}{4\lambda}}
\ee
is the Fourier transform of~\eqref{Gaussianpq}.

To perform the calculation~\eqref{Omega0}, it is useful to express $\hat{q}$ and $\hat{p}$ in terms of the creation and annihalation operators.
\be
\hat{q}=\frac{a+a^\dagger}{\sqrt{2}} \hspace{5mm} \hat{p}=\frac{a-a^\dagger}{i\sqrt{2}}
\ee
We find:
\be 
e^{-i y \hat{q}-i x \hat{p}}=e^{-\frac{a}{\sqrt{2}}(x+i y)-\frac{a^\dagger}{\sqrt{2}}(-x+i y)}=e^{a^\dagger \bar{t}-a t}
\ee
where $t=\frac{x+ i y}{\sqrt{2}}$. We now leverage the Baker-Campbell-Hausdorff formula:
\be
e^{a^\dagger \bar t-a t}=e^{a^\dagger \bar t} e^{- a t} e^{\frac{1}{2}[a^\dagger \bar t, a t]}=e^{-\frac{1}{2}|t|^2} e^{a^\dagger \bar t} e^{- a t}
\ee
We aim to express the integral in \eqref{Omega0} in terms of the complex variable $t$. To do so, we use the fact that
\be
\int \de x \de y= 2 \int \de^2t
\ee
We can express \eqref{Omega0} as follows:
\be 
\rho_{nm}=\bra n\hat\Omega(f)\ket m=\frac{1}{\pi}\int \de^2t \,  e^{-\frac{|t|^2}{2\lambda}} e^{-\frac{1}{2}|t|^2} \bra n e^{a^\dagger \bar t} e^{- a t}\ket m 
\ee
This is the same expression we find in Eq.~\eqref{Omegaf} for $s=0$, see \ref{App:proofofprop2} for the complete calculation, here let us write the result we are interested in. Let us define
\be
\alpha=\left(\frac{1}2+\frac{1}{2\lambda}\right)=\frac{1+\lambda}{2\lambda}
\ee
The result is
\be
\rho_{nm}=\rho_{nn}=\frac1\alpha\left(1-\frac1\alpha\right)^n=\frac{2\lambda}{1+\lambda}\left(\frac{1-\lambda}{1+\lambda}\right)^n 
\ee
Only the diagonal terms survive. Furthermore, since $\sum_{n=0}^\infty \left(1-\frac1\alpha\right)^n=\alpha$ we see that also in this case the trace of $\rho=1$.\\
To complete the proof we have to analyze what happens for different values of $\lambda$. The trace of this operator is always 1, but the eigenvalues are all positive  only for $\lambda<1$, and likewise  $\rho_{00}<1$ for the same range of $\lambda$. The case for $\lambda=1$ can be treated separately. For $n\neq 0$ then $\rho_{nn}=0$, while $\rho_{00}=1$. In this case the map associates to the Gaussian a \emph{pure} state: the eigenvector of the number operator corresponding to the zero eigenvalue,  (ground state for the harmonic oscillator). Only in this case the Weyl map associates a pure state. For $\lambda<1$ the Weyl maps associates a mixed state. For $\lambda>1$ the map still associates a Hermitean operator, but it cannot be considered a possible quantum state.

\subsection{Weyl quantization and Gaussianification} \label{App:proofth1}
Let us start with a probability distribution $\rho(q,p)$. Its Weyl-symmetric quantization, i.e.\ $\hat{\Omega}(f)$, is diagonal in the Fock basis (see~\cref{App:proofofprop2}). This is an important property, as it implies that all terms involving the action of either annihilation or creation operators on $\hat{\Omega}(f)$ vanish: 
\be 
\tr(a^n \hat{\Omega}(f))=\tr((a^\dagger)^n \hat{\Omega}(f))=0 \label{zero} 
\ee 
This is a remarkable result. In fact, recalling that 
\be 
\hat{q}=\frac{a + a^\dagger}{\sqrt{2}} 
 \ , \ \hat{p}=\frac{a - a^\dagger}{i\sqrt{2}} \ , \ \{\hat{q},\hat{p}\}=\frac{a^2-(a^\dagger)^2}{i} 
\ee 
we can exploit \eqref{zero}, yielding: 
\ba 
\tr(\hat{q}\hat{\Omega}(f))&=&\tr(\hat{p}\hat{\Omega}(f))=\tr(\{\hat{q},\hat{p}\}\hat{\Omega}(f))=0\\
\tr(\hat{q}^2\hat{\Omega}(f))&=&\tr(\hat{p}^2\hat{\Omega}(f))\label{eq2} 
\ea
Define $\Delta\coloneqq\tr(\hat{p}^2\hat{\Omega}(f))$. We have that $\hat{\Omega}(f)$ has the following covariance matrix
\be
V(\rho)=\begin{pmatrix}
    \Delta &0\\
    0&\Delta
\end{pmatrix}
\ee
Imposing the positivity condition of the Gaussification, i.e.\ Eq.~\eqref{eq:positivity} in the main text, we find the condition $\Delta\ge\frac{1}{2}$. We can dequantize the condition and see how it relates to the properties of the classical probability distribution. Recalling that we can dequantize symmetric polynomials of $\hat{q},\hat{p}$, given the symmetric quantization, we have the correspondence $\Delta\mapsto \frac{1}{2}(\Delta_q^2+\Delta_p^2)$, where $\Delta_q,\Delta_p$ are classical variances. We used the fact that the probability distribution is centered in the origin and has expectations of position and momentum zero. The condition  $\Delta_q^2+\Delta_p^2\ge 1$ is necessary for the positivity of the quantized probability distribution. This is due to the fact that non-symmetric probability distribution in $q,p$ are mapped to symmetric operators in $\hat{q},\hat{p}$, cfr.\ Eq.~\eqref{eq2}. 
The further requirement that the classical variances are equal to each other $\Delta_q=\Delta p$ recovers $\Delta_q\Delta_p\ge\frac{1}{2}$ as a necessary condition for positivity. If this is not the case, we can only assert that if $\hat{\Omega}(f)$ is non-positive, then necessarily $Delta_q\Delta_p<\frac{1}{2}$ (due to the fact that $\Delta_q\Delta_p\le\frac{1}{2}(\Delta_q^2+\Delta_p^2)$). 

\section{$s$-ordered quantization}
\subsection{Quantizer and dequantizer}\label{app:dequantizer}
Let us introduce the quantizer, which describes arbitrary orderings with respect to the creation and annihilation operators. These types of orderings were first introduced in~\cite{Cahill:1969it} (for a more modern analysis, see also~\cite{Manko:2004syv, Lizzi:2014pwa}). The quantizer can be expressed as follows:
\be
\hat D_s(z)=\int \de^2\xi\, e^{\xi(a^\dag -\bar z)-\bar\xi(a-z)}e^{\frac{s}2 |\xi|^2}
=\int \de^2\xi\, e^{\xi a^\dag}e^{-\bar\xi a}e^{-\frac{1-s}2 |\xi|^2}e^{-\xi\bar z+\bar\xi z} \label{quantz}
\ee
where in the last step we used the Baker-Campbell-Hausdorff formula.\\
The dequantizer exchanges $s\leftrightarrow -s$:
\be
\hat U_s(z)=\hat D_{-s}(z)
\ee 
It is now clear that the quantization map expressed in terms of the Fourier transform of the classic function $f(z)$ is:
\be
\hat\Omega_s(f)=\int \de^2\xi\, e^{\xi a^\dag}e^{-\bar\xi a}e^{-\frac{1-s}2 |\xi|^2}\tilde f(\xi)
\label{incorniciato}
\ee
The different orderings depend on $s$, which is why \eqref{incorniciato} is referred to as the $s$-ordered map. To illustrate the different orderings, let us present the following example:
\be\label{eq:quantizersec}
\hat\Omega_s(z\bar z)=\frac{1+s}2 a^\dag a + \frac{1-s}2 a a^\dag
\ee
we see that for $s=0$ we have the symmetric ordering, while for $s=1$  (resp.~$s=-1$) we normal (resp. antinormal) ordering.
 
\subsection{Cahill-Glauber quantization} \label{App:proofofprop2}
To determine the quantization of the Gaussian~\eqref{eq:complexgaussian}, we need to evaluate:
\be
\rho_{nm}=\bra n\hat\Omega_s(f)\ket m=\int \de^2\xi\,\bra n e^{\xi a^\dagger}e^{-\bar\xi a}e^{-\frac{1-s}2 |\xi|^2}\tilde \rho_\lambda(\xi) \ket m \label{Omegaf}
\ee
where
\be
\tilde\rho_\lambda(\xi)=\frac1{\pi}e^{-\frac{|\xi|^2}{2\lambda}} \label{transform}
\ee
is the Fourier transform of~\eqref{eq:complexgaussian}.\\
To calculate $\rho_{nm}$ we need:
\be
e^{-\bar\xi a}\ket m=\sum_{i=0}^m (-1)^i\frac{\bar\xi^i}{i!}\sqrt{\frac{m!}{(m-i)!}}\ket{m-i}
\ee
from which we obtain 
\begin{align}
\bra n e^{\xi a^\dag}e^{-\bar\xi a}\ket m=&\sum_{i=0}^m\sum_{j=0}^n \bra{n-j} (-1)^i\frac{\bar\xi^i}{i!}\frac{\xi^j}{j!}\sqrt{\frac{m!}{(m-i)!}}
\sqrt{\frac{n!}{(n-j)!}}\ket{m-i}\nonumber\\
=&\sum_{i=0}^m\sum_{j=0}^n  (-1)^i\frac{\bar\xi^i}{i!}\frac{\xi^j}{j!}\sqrt{\frac{m!}{(m-i)!}}
\sqrt{\frac{n!}{(n-j)!}}\delta_{n-j,m-i}
\end{align}
Assume for the time being that $n\geq m$ and set $n=m+k$. Then, since $\delta_{n-j,m-i}=\delta_{j,k+i}$, we have
\be
\bra{m+k} e^{\xi a^\dag}e^{-\bar\xi a}\ket m=\sqrt{m!(m+k)!}\sum_{i=0}^m \frac{(-1)^i}{i!} |\xi|^{2i} \bar\xi^k \frac1{(i+k)!}
\frac1{(m-i)!}
\ee
We can immediately anticipate that, setting $\xi=\frac r{\sqrt 2}e^{i\varphi}$, since $\tilde f(\xi)$ depends only on $r$, the angular integration of~\eqref{Omegaf} gives a factor of $2\pi$ and imposes $k=0$,  We have reached the important result that the operator $\hat\rho$ is diagonal in the discrete basis.
Hence
\be
\bra n e^{\xi a^\dag}e^{-\bar\xi a}\ket m=\delta_{m,n}\sum_{i=0}^n \frac{(-1)^i}{i!} |\xi|^{2i} \frac{n!}{i!(n-i)!} \label{off diagonal}
\ee
We can now calculate the full fledged integral:
\begin{align}
\rho_{nn}=&\sum_{i=0}^n\int\de^2\xi\, \frac{(-1)^i}{i!} |\xi|^{2i} \frac{n!}{i!(n-i)!} e^{-\frac{1-s}2 |\xi|^2}
\frac{e^{-\frac{|\xi|^2}{2\lambda}}}{\pi}\nonumber\\
=&\sum_{i=0}^n 2 \pi \int_0^\infty \frac{r}{2 \pi}\de r\, \frac{(-1)^i}{i!} r^{2i} (2)^{-i} \frac{n!}{i!(n-i)!} e^{-\left(\frac{1-s}4+\frac{1}{4\lambda}\right) r^2}
\end{align}
defining now 
\be
\alpha=\left(\frac{1-s}2+\frac{1}{2\lambda}\right)=\frac{1+(1-s)\lambda}{2\lambda}
\ee
 and using the fact that
\be
\int_0^\infty \de r\, r^{2i+1} e^{-\frac{\alpha r^2}2}=\frac{2^i i!}{\alpha^{i+1}}
\ee
we have
\begin{align}
\rho_{nn}=&\sum_{i=0}^n \frac{n!}{i!(n-i)!} \frac{(-1)^i}{\alpha^{i+1}}\nonumber\\
=&\frac1\alpha\left(1-\frac1\alpha\right)^n\nonumber\\
=&\frac{2\lambda}{1+(1-s)\lambda}\left(\frac{1-(1+s)\lambda}{1+(1-s)\lambda}\right)^n \label{result}
\end{align}
Since
$\sum_{n=0}^\infty \left(1-\frac1\alpha\right)^n=\alpha$ we see that also in this case the trace of $\rho=1$, which is a consequence of the fact that the quantization map is an isometry between Hilbert spaces.
\subsection{Temperature shift} \label{App:temperature}
Specifically, we compare the Fourier transform of the Gaussian in \eqref{eq:complexgaussian} with that of the Wigner function $e^{-\frac{2|z|^2}{2\beta+1}}$. Both expressions can be derived from equation \eqref{transform}. Additionally, we must recall that the quantizer introduces a factor $e^{\frac{s}{2}|\xi|^2}$ in the Fourier transform of the Gaussian (see Eq. \eqref{s quantization}). Thus, we need to compare $\frac{1}{\pi}e^{-\frac{|\xi|^2}{2\lambda}}e^{\frac{s}{2}|\xi|^2}$ with $\frac{1}{\pi}e^{-\frac{|\xi|^2 (2\beta+1)}{2}}$. From this comparison, we obtain:
\be
2\beta+1= \frac{1}{\lambda}-s \rightarrow \beta=\frac{1}{2}(\frac{1}{\lambda}-s-1)
\ee
Requiring $\beta\ge0$, we have $\lambda\le(1+s)^{-1}=\lambda_c(s)$. Similarly, for $\beta<0$, we find $\lambda>\lambda_c(s)$.
\subsection{s-ordering quantization and Gaussianification}\label{App:proofth3}
Let $\hat{\Omega}_s(f)$ be the quantized version of a classical probability density $\rho(z)$ with respect to the ordering of Eq.~\eqref{sordering} in the main text. Similarly to the Weyl symmetric ordering, we have that $\hat{\Omega}_s(f)$ is diagonal in the Fock basis. Hence Eq.~\eqref{zero} and Eq.~\eqref{eq2} still hold. Defining 
\be
\Delta\coloneqq\tr(\hat{q}^2\hat{\Omega}_s(f))=\tr(a^2\hat{\Omega}_s(f))+\tr((a^{\dagger})^2\hat{\Omega}_s(f))+\tr(\{a,a^{\dag}\}\hat{\Omega}_s(f))=\tr(\{a,a^{\dag}\}\hat{\Omega}_s(f)) 
\ee 
where the last equality comes from Eq.~\eqref{zero}. \\
We have that a necessary condition for positivity is $\Delta\ge\frac{1}{2}$. This holds following the same passages as in~\cref{App:proofth1}. To recover the condition on the classical probability distribution, we can invert the condition as follows. First we write
\be
\hat{\Omega}_s(f)=\int\de^2z\rho(z)\hat{D}_s(z)
\ee
where 
\be
\hat{D}_s\coloneqq \int\de^2\xi e^{-\xi \bar{z}+z\bar{\xi}}e^{\xi a^{\dag}}e^{-\bar{\xi}a}e^{-\frac{1-s}{2}|\xi|^2}
\ee
Then express $\Delta$ as 
\be
\Delta=\int\de^2z\rho(z)\tr(\{a,a^{\dag}\}\hat{D}_s)= \int\de^2z\rho(z)\tr(\{a,a^{\dag}\}\hat{\Omega}^{-1}_{-s})\label{eq3}
\ee
where we have identified $\hat{D}_s=\hat{\Omega}_{-s}^{-1}$, see Section~\ref{app:dequantizer}. We can rewrite
\ba
\{a,a^{\dag}\}&=&\frac{1-s}{2}a^{\dag}a+\frac{1+s}{2}aa^{\dag}+\frac{1-s}{2}aa^{\dag}+\frac{1+s}{2}a^{\dag}a\\&=&
2\left(\frac{1-s}{2}a^{\dag}a+\frac{1+s}{2}aa^{\dag}\right)+\frac{1-s}{2}-\frac{1+s}{2}\\
&=&2\left(\frac{1-s}{2}a^{\dag}a+\frac{1+s}{2}aa^{\dag}\right)-s
\ea
Applying $\hat{\Omega}^{-1}_{-s}$, and using Eq.~\eqref{eq:quantizersec}, we thus have
\be
\hat{\Omega}^{-1}_{-s}(\{a,a^{\dag}\})=2z\bar{z}-s
\ee
Substituting back into Eq.~\eqref{eq3}, we end up with the condition on the classical probability distribution
\be
2\int\de^2z\rho(z)|z|^2\ge\frac{1}{2}+s\int\de^2z\rho(z)
\ee
Going back to the $p,q$ variables we arrive to
\be
\Delta_{q}^2+\Delta_{p}^2\ge(1+s)
\ee


\begin{thebibliography}{10}

\bibitem{Weyl}
H.~Weyl, {\em The Theory of Groups and Quantum Mechanics}.
\newblock Dover, New York, 1931.

\bibitem{suleymanov2019wigner}
M.~Suleymanov and M.~Zubkov, ``Wigner--Weyl formalism and the propagator of
  Wilson fermions in the presence of varying external electromagnetic field,''
  {\em Nuclear Physics B} {\bfseries 938} (2019) 171--199.

\bibitem{Zampini:2005rx}
A.~Zampini, ``{Applications of the Weyl-Wigner formalism to noncommutative
  geometry},'' other thesis, Università di Napoli \textit{Federico II}, 5,
  2005.

\bibitem{tatarskiui1983wigner}
V.~Tatarskiĭ, ``The Wigner representation of quantum mechanics,'' {\em Soviet
  Physics Uspekhi} {\bfseries 26} no.~4, (1983) 311.

\bibitem{Wigner}
E.~P. Wigner, ``On the Quantum Correction For Thermodynamic Equilibrium,''
  \href{http://dx.doi.org/10.1103/PhysRev.40.749}{{\em Phys. Rev.} {\bfseries
  40} (1932) 749}.

\bibitem{Ferrie_2008}
C.~Ferrie and J.~Emerson, ``Frame representations of quantum mechanics and the
  necessity of negativity in quasi-probability representations,''
  \href{http://dx.doi.org/10.1088/1751-8113/41/35/352001}{{\em Journal of
  Physics A: Mathematical and Theoretical} {\bfseries 41} no.~35, (July, 2008)
  352001}. \url{http://dx.doi.org/10.1088/1751-8113/41/35/352001}.

\bibitem{Ferrie_2011}
C.~Ferrie, ``Quasi-probability representations of quantum theory with
  applications to quantum information science,''
  \href{http://dx.doi.org/10.1088/0034-4885/74/11/116001}{{\em Reports on
  Progress in Physics} {\bfseries 74} no.~11, (Oct., 2011) 116001}.
  \url{http://dx.doi.org/10.1088/0034-4885/74/11/116001}.

\bibitem{Veitch_2012}
V.~Veitch, C.~Ferrie, D.~Gross, and J.~Emerson, ``Negative quasi-probability as
  a resource for quantum computation,''
  \href{http://dx.doi.org/10.1088/1367-2630/14/11/113011}{{\em New Journal of
  Physics} {\bfseries 14} no.~11, (Nov., 2012) 113011}.
  \url{http://dx.doi.org/10.1088/1367-2630/14/11/113011}.

\bibitem{Mari_2012}
A.~Mari and J.~Eisert, ``Positive Wigner Functions Render Classical Simulation
  of Quantum Computation Efficient,''
  \href{http://dx.doi.org/10.1103/physrevlett.109.230503}{{\em Physical Review
  Letters} {\bfseries 109} no.~23, (Dec., 2012) 230503}.
  \url{http://dx.doi.org/10.1103/PhysRevLett.109.230503}.

\bibitem{Gard_2015}
B.~T. Gard, K.~R. Motes, J.~P. Olson, P.~P. Rohde, and J.~P. Dowling, {\em An
  Introduction to Boson-Sampling},
  \href{http://dx.doi.org/10.1142/9789814678704_0008}{p.~167–192}.
\newblock WORLD SCIENTIFIC, June, 2015.
\newblock \url{http://dx.doi.org/10.1142/9789814678704_0008}.

\bibitem{doi:10.1126/science.abe8770}
H.-S. Zhong, H.~Wang, Y.-H. Deng, M.-C. Chen, L.-C. Peng, Y.-H. Luo, J.~Qin,
  D.~Wu, X.~Ding, Y.~Hu, P.~Hu, X.-Y. Yang, W.-J. Zhang, H.~Li, Y.~Li,
  X.~Jiang, L.~Gan, G.~Yang, L.~You, Z.~Wang, L.~Li, N.-L. Liu, C.-Y. Lu, and
  J.-W. Pan, ``Quantum computational advantage using photons,''
  \href{http://dx.doi.org/10.1126/science.abe8770}{{\em Science} {\bfseries
  370} no.~6523, (2020) 1460--1463},
  \href{http://arxiv.org/abs/https://www.science.org/doi/pdf/10.1126/science.abe8770}{{\ttfamily
  https://www.science.org/doi/pdf/10.1126/science.abe8770}}.
  \url{https://www.science.org/doi/abs/10.1126/science.abe8770}.

\bibitem{Lund_2017}
A.~P. Lund, M.~J. Bremner, and T.~C. Ralph, ``Quantum sampling problems,
  BosonSampling and quantum supremacy,''
  \href{http://dx.doi.org/10.1038/s41534-017-0018-2}{{\em npj Quantum
  Information} {\bfseries 3} no.~1, (Apr., 2017) 15}.
  \url{http://dx.doi.org/10.1038/s41534-017-0018-2}.

\bibitem{Cahill:1969it}
K.~E. Cahill and R.~J. Glauber, ``{Ordered expansions in boson amplitude
  operators},'' \href{http://dx.doi.org/10.1103/PhysRev.177.1857}{{\em Phys.
  Rev.} {\bfseries 177} (1969) 1857--1881}.

\bibitem{PhysRevA.90.013810}
C.~Hughes, M.~G. Genoni, T.~Tufarelli, M.~G.~A. Paris, and M.~S. Kim, ``Quantum
  non-Gaussianity witnesses in phase space,''
  \href{http://dx.doi.org/10.1103/PhysRevA.90.013810}{{\em Phys. Rev. A}
  {\bfseries 90} (Jul, 2014) 013810}.
  \url{https://link.aps.org/doi/10.1103/PhysRevA.90.013810}.

\bibitem{ivan2012measure}
J.~S. Ivan, M.~S. Kumar, and R.~Simon, ``A measure of non-Gaussianity for
  quantum states,'' {\em Quantum information processing} {\bfseries 11} no.~3,
  (2012) 853--872.

\bibitem{tan2020negativity}
K.~C. Tan, S.~Choi, and H.~Jeong, ``Negativity of quasiprobability
  distributions as a measure of nonclassicality,'' {\em Physical review
  letters} {\bfseries 124} no.~11, (2020) 110404.

\bibitem{de2019measuring}
S.~De~Bievre, D.~B. Horoshko, G.~Patera, and M.~I. Kolobov, ``Measuring
  nonclassicality of bosonic field quantum states via operator ordering
  sensitivity,'' {\em Physical Review Letters} {\bfseries 122} no.~8, (2019)
  080402.

\bibitem{berra2020coherent}
J.~Berra-Montiel and A.~Molgado, ``Coherent representation of fields and
  deformation quantization,'' {\em International Journal of Geometric Methods
  in Modern Physics} {\bfseries 17} no.~11, (2020) 2050166.

\bibitem{berra2020star}
J.~Berra-Montiel, ``Star product representation of coherent state path
  integrals,'' {\em The European Physical Journal Plus} {\bfseries 135} no.~11,
  (2020) 906.

\bibitem{Lee1995}
H.-W. Lee, ``Theory and application of the quantum phase-space distribution
  functions,'' \href{http://dx.doi.org/10.1016/0370-1573(95)00007-4}{{\em
  Physics Reports} {\bfseries 259} no.~3, (1995) 147--211}.
  \url{https://doi.org/10.1016/0370-1573(95)00007-4}.

\bibitem{Kempf:1994su}
A.~Kempf, G.~Mangano, and R.~B. Mann, ``{Hilbert space representation of the
  minimal length uncertainty relation},''
  \href{http://dx.doi.org/10.1103/PhysRevD.52.1108}{{\em Phys. Rev. D}
  {\bfseries 52} (1995) 1108--1118},
  \href{http://arxiv.org/abs/hep-th/9412167}{{\ttfamily arXiv:hep-th/9412167}}.

\bibitem{Brandenberger:2002nq}
R.~Brandenberger and P.-M. Ho, ``{Noncommutative space-time, stringy space-time
  uncertainty principle, and density fluctuations},''
  \href{http://dx.doi.org/10.1103/PhysRevD.66.023517}{{\em Phys. Rev. D}
  {\bfseries 66} (2002) 023517},
  \href{http://arxiv.org/abs/hep-th/0203119}{{\ttfamily arXiv:hep-th/0203119}}.

\bibitem{Chamseddine_2010}
A.~Chamseddine and A.~Connes, ``Noncommutative geometry as a framework for
  unification of all fundamental interactions including gravity. Part I.,''
  \href{http://dx.doi.org/10.1002/prop.201000069}{{\em Fortschritte der Physik}
  {\bfseries 58} no.~6, (May, 2010) 553–600}.
  \url{http://dx.doi.org/10.1002/prop.201000069}.

\bibitem{Devastato:2019grb}
A.~Devastato, M.~Kurkov, and F.~Lizzi, ``{Spectral Noncommutative Geometry,
  Standard Model and all that},''
  \href{http://dx.doi.org/10.1142/S0217751X19300102}{{\em Int. J. Mod. Phys. A}
  {\bfseries 34} no.~19, (2019) 1930010},
  \href{http://arxiv.org/abs/1906.09583}{{\ttfamily arXiv:1906.09583
  [hep-th]}}.

\bibitem{Huggett:2020kok}
N.~Huggett, F.~Lizzi, and T.~Menon, ``{Missing the point in noncommutative
  geometry},'' \href{http://dx.doi.org/10.1007/s11229-020-02998-1}{{\em
  Synthese} {\bfseries 199} no.~1-2, (2021) 4695--4728},
  \href{http://arxiv.org/abs/2006.13035}{{\ttfamily arXiv:2006.13035
  [physics.hist-ph]}}.

\bibitem{WANG_2007}
X.~Wang, T.~Hiroshima, A.~Tomita, and M.~Hayashi, ``Quantum information with
  Gaussian states,''
  \href{http://dx.doi.org/10.1016/j.physrep.2007.04.005}{{\em Physics Reports}
  {\bfseries 448} no.~1–4, (Aug., 2007) 1–111}.
  \url{http://dx.doi.org/10.1016/j.physrep.2007.04.005}.

\bibitem{Weedbrook_2012}
C.~Weedbrook, S.~Pirandola, R.~García-Patrón, N.~J. Cerf, T.~C. Ralph, J.~H.
  Shapiro, and S.~Lloyd, ``Gaussian quantum information,''
  \href{http://dx.doi.org/10.1103/revmodphys.84.621}{{\em Reviews of Modern
  Physics} {\bfseries 84} no.~2, (May, 2012) 621–669}.
  \url{http://dx.doi.org/10.1103/RevModPhys.84.621}.

\bibitem{adesso2014continuous}
G.~Adesso, S.~Ragy, and A.~R. Lee, ``Continuous variable quantum information:
  Gaussian states and beyond,'' {\em Open Systems \& Information Dynamics}
  {\bfseries 21} no.~01n02, (2014) 1440001.

\bibitem{genoni2016conditional}
M.~G. Genoni, L.~Lami, and A.~Serafini, ``Conditional and unconditional
  Gaussian quantum dynamics,'' {\em Contemporary Physics} {\bfseries 57} no.~3,
  (2016) 331--349.

\bibitem{mele2024learningquantumstatescontinuous}
F.~A. Mele, A.~A. Mele, L.~Bittel, J.~Eisert, V.~Giovannetti, L.~Lami,
  L.~Leone, and S.~F.~E. Oliviero, ``Learning quantum states of continuous
  variable systems,'' 2024.
\newblock \url{https://arxiv.org/abs/2405.01431}.

\bibitem{Moyal}
J.~E. Moyal, ``Quantum mechanics as a statistical theory,''
  \href{http://dx.doi.org/10.1017/S0305004100000487}{{\em Proc. Cambr. Phil.
  Soc.} {\bfseries 45} (1949) 99}.

\bibitem{Groenewold:1946kp}
H.~J. Groenewold, ``{On the Principles of elementary quantum mechanics},''
  \href{http://dx.doi.org/10.1016/S0031-8914(46)80059-4}{{\em Physica}
  {\bfseries 12} (1946) 405--460}.

\bibitem{Curtright:2011vw}
T.~L. Curtright and C.~K. Zachos, ``{Quantum Mechanics in Phase Space},''
  \href{http://dx.doi.org/10.1142/S2251158X12000069}{{\em Asia Pac. Phys.
  Newslett.} {\bfseries 1} (2012) 37--46},
  \href{http://arxiv.org/abs/1104.5269}{{\ttfamily arXiv:1104.5269
  [physics.hist-ph]}}.

\bibitem{derezinski2020}
J.~Derezi{\'n}ski and M.~Karczmarczyk, ``Quantization of gaussians,'' {\em
  Analysis as a Tool in Mathematical Physics: In Memory of Boris Pavlov} (2020)
  277--304, \href{http://arxiv.org/abs/1701.07297}{{\ttfamily arXiv:1701.07297
  [math-ph]}}. \url{https://arxiv.org/abs/1701.07297}.

\bibitem{cahen2023complexweylsymbolsmetaplectic}
B.~Cahen, ``Complex Weyl symbols of metaplectic operators: an elementary
  approach,'' 2023.
\newblock \url{https://arxiv.org/abs/2306.12947}.

\bibitem{BUCCO}
A.~Serafini, {\em Quantum continuous variables: A primer of theoretical
  methods}.
\newblock CRC Press, Taylor \& Francis Group, Boca Raton, USA, 2017.

\bibitem{Manko:2004syv}
V.~I. Man'ko, G.~Marmo, and P.~Vitale, ``{Phase space distributions and a
  duality symmetry for star products},''
  \href{http://dx.doi.org/10.1016/j.physleta.2004.11.027}{{\em Phys. Lett. A}
  {\bfseries 334} (2005) 1},
  \href{http://arxiv.org/abs/hep-th/0407131}{{\ttfamily arXiv:hep-th/0407131}}.

\bibitem{Lizzi:2014pwa}
F.~Lizzi and P.~Vitale, ``{Matrix Bases for Star Products: a Review},''
  \href{http://dx.doi.org/10.3842/SIGMA.2014.086}{{\em SIGMA} {\bfseries 10}
  (2014) 086}, \href{http://arxiv.org/abs/1403.0808}{{\ttfamily arXiv:1403.0808
  [hep-th]}}.

\end{thebibliography}

\providecommand{\href}[2]{#2}\begingroup\raggedright\endgroup

\end{document}